\documentclass[12pt]{article} 
\usepackage{amsmath,amssymb,amscd}
\usepackage{graphicx,epsfig}

\begin{document} 
\begin{center}
       {\bf \Large   The quark-gluon medium\\ (micro- and macro-QCD)}               \\

\vspace{2mm}

{\bf I.M. Dremin\footnote{Email: dremin@lpi.ru}} \\

\vspace{2mm}

{\it Lebedev Physical Institute, Moscow 119991, Russia}\\

\end{center}

\begin{abstract} 
The properties of the quark-gluon medium observed in high energy 
nucleus-nucleus collisions are discussed. The main experimental facts 
about these collisions are briefly described and compared with data
about proton-proton collisions. Both microscopic and macroscopic approaches 
to their description are reviewed. The chromodynamics of the quark-gluon medium
at high energies is mainly considered. The energy loss of partons moving in 
this medium is treated. The principal conclusion is that the medium 
possesses some collective properties which are crucial for understanding
the experimental observations.
\end{abstract}


\section{Introduction}

The topic of {\it the quark-gluon medium} is so widely spread nowadays that 
it can not be squeezed into a single lecture. That is why I mostly concentrate  
on general ideas about the evolution of the quark-gluon medium in high energy 
heavy ion collisions and on those properties of the medium which are revealed 
by the energy losses of partons moving in it. They are described by the 
chromodynamics of the quark-gluon medium which is of the main concern in this 
lecture. It ascribes quarks and gluons as elementary objects (partons) 
responsible for the interaction. The goal is to confront experimental data 
and theoretical ideas about the properties of the quark-gluon medium
formed in high energy collisions of hadrons and nuclei (for recent review
see \cite{dl10}).

General "consensus" is achieved concerning the types of processes observed in 
pp-collisions and their description. They are split into the two groups:
SOFT (low $p_T$) processes, where one uses phenomenology, notion of clusters, 
multiperipheral models, BFKL equations; and
HARD (large $p_T$) processes, where one considers perturbative QCD, 
jets production and their evolution, DGLAP equations. 

Highly coherent parton configurations and strong internal fields become 
especially important for the matter in collision. 
By colliding two heavy nuclei at ultrarelativistic energies one expects to get
a hot and dense internally colored medium. It should exhibit some collective  
properties not seen at static conditions. 

\section{The main experimental findings}

Nowadays, high energy experimental data are obtained at the following 
accelerators: Tevatron ($p\bar p$, $\; \sqrt s \leq $ 1.96 TeV);
RHIC (pp and ions (up to Au+Au), $\; \sqrt s \leq $ 200A GeV);
LHC (pp, $\; \sqrt s = $ 0.9; 2.36; 7; planned 14 TeV; and Pb+Pb, $\; 
\sqrt s = $ 2.76A TeV; planned 5.5A TeV).

The very first question addressed is the difference in AA and pp processes at
high energies. The following characteristics were studied:

 1.Ratio of particle spectra in AA and pp
\begin{equation}
R_{AA}(y, p_T;b)=\frac {d^2N_{AA}/dydp_T}{N_{part}d^2N_{pp}/dydp_T} 
\end{equation}
shows that $R_{AA}\approx 0.2$ at high $p_T>4$ GeV which indicates jet 
quenching (both at RHIC and LHC) and leads to conclusion that AA are not reducible to pp.

2. Correlations: two-, three-particle; jets; BE-HBT; ridge;
double-humped distributions... 
They reveal, e.g., collective flows: $v_2$ - elliptic (azimuthal) flow induced by the 
liquid pressure
\begin{equation}
E\frac {d^3N}{d^3p}=\frac {1}{2\pi }\frac {d^2N}{dyp_Tdp_T}(1+
\sum _{n=1}^{\infty }2v_n\cos[n(\phi -\Psi _r)])
\end{equation}

3. Enhancement: hadrocomposition changes (e.g., strangeness, 
p/$\pi $, quarkonia ...)

These facts lead to CONCLUSION: The quark-gluon medium is formed (plasma,
liquid).

Many important features were observed in 7 TeV pp-collisions at LHC.
For example, particle density in pp (already at 2.36 TeV) becomes comparable to that of 
AA at RHIC; the size of the interaction region at high multiplicities becomes
comparable to that of AA at RHIC (as found from Bose-Einstein correlations),
and, most spectacular, ridge was observed in high multiplicity pp at 7 TeV.
It raises the question whether this is the signature of the quark-gluon medium
formed even in pp. No definite answer was found yet.

\section{From partons and fields to the quark-gluon medium}

The evolution of quark-gluon fields at the initial stages of collisions is
described by microscopic QCD. Later, due to some instabilities the collective
properties of this system develop and a quark-gluon matter is formed. At this
stage its properties (in particular, propagation of partons through it) should
be describable both by micro- and macro-QCD.

At the very beginning, the two Lorentz-contracted sheets with transverse
partonic fields (called the Color Glass Condensate - CGC) collide. Soon
afterwards they transform into a system of longitudinal fields in 
color flux tubes localized in the transverse plane and stretching between 
the valence color degrees of freedom. It is called Glasma.
This evolution is described by the JIMWLK-equation \cite{jim} obtained from
micro-QCD. It preceeds the next stage where the matter with collective 
properties is formed (possibly, after some instabilities). It is named the
Quark-Gluon Plasma (QGP). At the final stage it hadronizes producing final
particles.

The mechanical and thermodynamical properties of QGP are studied by comparison
of experimental data with theoretical results obtained in the framework
of the QCD-inspired hydrodynamics briefly discussed in part 4 of the lecture.

The energy loss of partons in the quark-gluon medium is the main source of
experimental information about its properties during the QGP stage. As in
electrodynamics, it may be separated in the two categories.

The loss due to the change of the velocity vector of a parton such as elastic 
scattering, bremsstrahlung and synchrotron radiation is usually
treated microscopically. All these phenomena result from the {\it short-range} 
response of the parton to the impact of the matter fields. Elastic scattering
does change the parton energy due to the recoil effect and deflects it and thus changes the energy flow in the
initial direction. At high energies it is less probable than the emission of
gluons (bremsstrahlung) during the nearby collisions with the matter
constituents. The medium structure imposes some conditions on the coherence
properties of the radiation process and on the effective radiation length
(in analogy to the Ter-Mikhaelyan and Landau-Pomeranchuk effects in
electrodynamics). In general, the intensity of the radiation depends on the 
relation on the path length of the parton in the medium, its mean free path 
(defined by the distance between the scattering centers and by cross sections) 
and the formation length of emitted radiation. 
The synchrotron radiation of gluons induced by the curvature
of parton trajectory in chromomagnetic fields may become important for
strong enough fields. Namely these processes are considered in part 5 of the
lecture.

Principally different source of energy losses is connected with the medium
polarization by the propagating parton. This is the collective response of the 
medium related to the non-perturbative {\it long-range} interconnection of its
constituents described by its chromopermittivity. It is treated macroscopically
by the in-medium QCD. The corresponding effects are analogous to the Cherenkov
radiation, the wake and the transition radiation. In these processes one can, 
in the first approximation, 
neglect the change of the velocity vector of a parton. The macroscopical 
aspect of the problem is often ignored and I will 
discuss it in more details in the part 6 of the lecture.

The total energy loss is the sum of the both micro- and macro-losses.

\section{Hydrodynamics (thermodynamical and mechanical properties of QGP)}

The thermodynamical and mechanical characteristics of QGP change during its
evolution. Spatial and temporal information is necessary to get them. The 
lattice calculations and ideas about CGC, Glasma, QGP provide some hints to
transition from confined state to deconfined quarks and gluons. The inelastic
collisions may lead to thermalization of the medium. Subsequently, it expands
and hydrodynamics may be applied to treat this stage of evolution. It is
described in many reviews (e.g., see \cite{sh1, hein}) and I give just a brief 
survey of it here. 

The medium is characterized by 6 independent variables. Those are the energy
density $e$, pressure $p$, baryon number $n_B$ and 3 components of the
velocity vector $u_{\mu }$. The energy-momentum tensor and baryon number
current are
\begin{equation}
T^{\mu \nu }(x)=(e(x)+p(x))u^{\mu }(x)u^{\nu }(x)-p(x)g^{\mu \nu },
\;\;\;  j_B^{\mu }(x)=n_B(x)u^{\mu }(x).     \label{tj} 
\end{equation} 
Six hydrodynamical equations of motion govern these variables. Those are five
non-linear partial differential equations obtained from local conservation laws
for energy, momentum and baryon number
\begin{equation}
\partial _{\mu } T^{\mu \nu }(x)=0 \;\; (\nu = 0, ..., 3);   
\;\;\;  \partial _{\mu }j_B^{\mu }(x)=0     \label{ctj} 
\end{equation} 
with an equation of state relating $p,\;e$ and $n_B$. It is usually chosen
to mimic the lattice QCD results by matching the states below and above the 
critical temperature (i.e., hadrons to a quark-gluon medium). This is 
rather indefinite element of the whole approach. Besides, the solutions of 
the non-linear equations in (3+1)-dimensions ask for initial conditions to
be defined and may be only obtained numerically with several external
parameters. The shape of initial hydro profile is a major source of
uncertainty. That is the origin of several conflicting results. The CGC 
and Glasma approaches provide some guesses to the initial conditions for the 
evolution of a thermalized QGP. However, the evolution of the Glasma into a 
thermalized QGP is not yet understood.

At least four
parameters are necessary to fix the initial conditions and freeze-out algorithm
which determines the transition from the hydrodynamical characteristics to
hadronic stage. Those are the initial time, energy (or entropy) density, 
baryon number  and the freeze-out temperature (or decoupling energy density)
$\tau _{eq}; \; s_{eq}; \; n_{B,eq}; \; e_{dec}$. They are fixed by further 
comparison with experimental data. Schematically, the correspondence of these
parameters and experimental characteristics can be represented as
$dN/dp_T - T - e_{dec}; \; dN/dy - (\tau s)_{eq}; \; 
p/\pi - n_{B,eq}/s_{eq}; \; (dN/dp_T)_p/(dN/dp_T)_{\pi } - \tau _{eq}$. 
Moreover, from the overlap geometry of the two colliding nuclei one should 
define the shape of the initial density profiles as well as assume the 
profiles of the initial longitudinal and transverse flows and final 
hadronization prescriptions. 

From solutions of the hydrodynamical equations one gets the transverse momentum 
spectra for various species of particles, the radial and elliptic flows, the
shape of the interaction region (as revealed by Bose-Einstein correlations 
and Hanburry Brown-Twiss interferometry). Their comparison with experimental 
data allows to determine the main thermodynamical, statistical 
and mechanical properties of the quark-gluon medium. 
Symbolically, their values found from 200 GeV data can be grouped as
$T_{eq}\approx 360$ MeV; $T_{cr}\approx 170$ MeV;
$T_{dec}\approx 120$ MeV; $\tau _{therm}=\tau _{eq}\approx 0.6<1$ fm; 
$\tau _{dec}\approx 7$ fm; 
$e_{th}\approx 25$ GeV/fm$^3$; $e_{cr}\approx 1$ GeV/fm$^3$;
$e_{dec}\approx 0.075$ GeV/fm$^3$; $s_{eq}\approx 110$ fm$^{-3}$;
$\eta /s\approx 0.1$ ($1/4\pi $ in AdS/CFT); $n_B<0.5$ fm$^{-3}$. 
They show rapid thermalization, high average initial energy density and quite 
long "lifetime" of the quark-gluon plasma before hadronization at rather low 
energy density and temperature predicted by lattice QCD. Complete thermalization 
in the time less than 1 fm is required to obtain the measured value of elliptic flow
and its centrality dependence which is very sensitive to any deviation from 
local thermal equilibrium of low $p_T$ particles. Collective excitations, 
resonances and inelastic collisions keep the system in thermal equilibrium. 
The good agreement of the data 
with ideal fluid dynamics points to a very small viscosity of QGP. Other
transport coefficients (shear, diffusion, heat conduction) are not important 
if the microscale defined by rescattering is much less than the macroscale 
related to medium expansion. Strong non-perturbative interaction
should be responsible for its behavior as an ideal liquid. Herefrom comes
the name of strong quark-gluon plasma (sQGP). In sQGP, there may exist, e.g., 
clusters \cite{yuk} and colored
bound states of massive quasiparticles \cite{szah} with heavier flavors. 
Resonance scattering on constituents of the quark-gluon medium can become 
important \cite{rapp}. All that 
would give rise to its collective response with new scales discussed above, 
high pressure, large chromopermittivity and long-range correlations necessary 
to explain the enhancement of strange partons production by non-local processes. 
The long "lifetime" supports approximate description of energy losses within 
an "infinite" medium.

Hydrodynamics is an actively developing field nowadays. Main characteristics
of low $p_T$ particles have been fitted in this approach albeit many factors
have to be taken into account such as thermodynamics, models of collective
flows, hadronization process, resonance decays, chemical composition, geometry 
of collisions etc. However a fully consistent hydrodynamical description has 
not yet been found. Some controversial conclusions appear from time to time 
about , e.g., energy (from SPS to LHC), rapidity and centrality dependence 
of elliptic flow, its absolute value, the transverse 
momentum dependence of various radii derived from HBT analysis, the chemical 
composition of some species (e.g., $\bar {p}/\pi $ ratio) etc. One may hope 
that they will be resolved in the near future within the same set of adjusted
parameters. That would provide deeper understanding of collective 
thermodynamical and mechanical properties of the bulk matter and its evolution.

\section{Parton energy loss in micro-QCD}

Experimentalists observe collimated groups of particles called jets created
in high energy collisions. There are different stages of jet formation:
\begin{itemize}
\item{Production of the initial high energy parton (described by pQCD).}
\item{The partonic stage of jet evolution in the medium \\
with account of {\it external} fields (described by DGLAP equations 
with account of the LPM-effect).}
\item{Hadronization - final particles production (using the parton
distribution functions known from experiment).} 
\end{itemize}

Two parameters are especially crucial for such description: 
${\hat q} = \langle p^2_\perp \rangle / \lambda $ and $L-$the medium size.
as well as their combination  $(\omega_c = \frac{1}{2} {\hat q} L^2)$.

The physical problem at hand is the computation of the leading contribution to 
the medium-induced energy loss $\Delta E$ of the parton produced inside the 
medium on its way from the hot dense fireball. Generically the energy loss 
$\Delta E$ depends on the medium thickness $L$ and the bulk characteristics of 
the medium and is described by some probability distribution 
${\cal P} (\Delta E)$.

In the general case the losses are determined by the following factors:

\begin{itemize}
\item{Probability of an elementary event leading to energy loss is characterized by the opacity $N=L/\lambda$, where $\lambda$ is the mean
free path of the parton in the medium under consideration. For the particle with integrated particle-medium interaction cross section $\sigma$ and
the medium density $\rho$ the mean free path can be estimated as $\lambda \sim 1/(\rho \sigma)$. }
\item{The intensity of the impact or scattering power of the medium is characterized by the transport coefficient
${\hat q} = \langle p^2_\perp \rangle / \lambda$, where $\langle p^2_\perp \rangle $ is the average transverse momentum squared that the
propagating particle gets from the elementary act of collision. In the thermal medium ${\hat q}=m^2_D/\lambda$ where $m_D$ is the Debye mass.}
\end{itemize}

At present there exists a general consensus, supported by extensive model calculations, that the radiative medium-induced energy loss 
is the dominant one, see e.g. \cite{Z07}.    Theoretical work on studying the properties of medium-induced gluon radiation was carried out by 
several groups \cite{BDMPS, GLV, W00, ASW, Z96, Z, Z99, HT, AMY}. The approaches of the above-listed groups 
differ in details of treating the relevant kinematics of the radiation and the 
details of describing the medium under consideration. 

Unfortunately, variations of $p_\perp $ per unit length, described by the 
parameter ${\hat q}$, differ sometimes by an order of magnitude in different 
models when compared with experiment. 

In the limit $\omega \ll \omega_c$ the medium acts coherently reducing the 
intensity of radiation while that of $\omega \gg \omega_c$ corresponds to incoherent scattering.
\begin{eqnarray}
 \left. \omega \frac{dI}{d\omega} \right \vert_{\omega \ll \omega_c} & 
 \thickapprox & \alpha_s \sqrt{\frac{{\hat q}L^2}{\omega}}, \nonumber \\
 \left. \omega \frac{dI}{d\omega } \right \vert_{\omega \gg \omega_c} & 
 \thickapprox & \alpha_s \frac{{\hat q}L^2}{\omega }.
\end{eqnarray}
In the soft limit $\omega \ll \omega_c$ one has for the average energy loss
\begin{equation}
\langle \Delta E \rangle \backsimeq \int_0^{\omega _c} d\omega \omega 
\frac{dI}{d\omega } \sim {\hat q} L^2.
\end{equation}
For some time one thought that the quadratic dependence on the medium thickness 
$L$ is a specifically non-Abelian effect. Recently, however, in \cite{SP09}, 
see also \cite{Z04}, it was shown that 
the effect is generic and is valid both in QED and QCD for the radiation 
of particle produced inside the medium at the initial piece of its trajectory.

Thus the finite length is crucial. The key question is therefore to choose, for 
describing the in-medium QCD cascade, the language allowing a credible 
space-time interpretation. As emphasized in \cite{JEWEL}, the only scheme 
allowing such interpretation is the old-style PYTHIA virtuality-ordered cascade 
where connection to the spatiotemporal pattern is achieved by calculating the 
lifetime $\tau$ of a virtual parton which, for a parton with the energy $E$ and 
virtuality $Q^2$ that has been created in the decay of its parent parton with 
the virtuality $Q^2_{\rm par}$, reads
\begin{equation}\label{formtime}
\tau = E \left( \frac{1}{Q^2}-\frac{1}{Q^2_{\rm par}} \right).
\end{equation}
Another important ingredient is the statement about the angular ordering. While
valid in vacuum, it can be modified for in-medium cascades.
It has been shown in \cite{LN10} that the properties of the cascade can be 
drastically changed if these problems are properly treated.

\section{The macroscopic approach to the quark-gluon medium}

Macroscopic collective properties of a medium may reveal themselves in its
mechanical motion as a whole (e.g., viscosity) described by hydrodynamics
or in its response to external color currents (e.g., chromopermittivity)
described by in-medium QCD. 

The medium itself can radiate in response to permutations. The
genuine role of the medium and its collective properties are most
clearly revealed by its polarization ${\bf P}$ due to the external current.
The macroscopic approach to the description of such collective properties is the most suitable one. The linear 
response of the medium to the electromagnetic field ${\bf E}$ is usually 
described as
\begin{equation}
{\bf P}= \frac {\epsilon - 1}{4\pi }{\bf E},     \label{pe}
\end{equation}
where $\epsilon $ is the dielectric (in macro-QED) or 
chromoelectric (in macro-QCD) permittivity. It is seen that the 
polarization can be quite strong for large values of $\epsilon $. The initial 
radiation process serves as a trigger for the collective response of the
medium initiated by the polarization. The well known examples are provided by
the Cherenkov radiation, the wake and the transition radiation.

What is typical for their description is the (approximate) constancy of the 
particle velocity vector used in the external current. In ultrarelativistic 
processes $(\gamma \gg 1$) the relative change of the velocity is much smaller 
than the relative energy loss because they are connected by the formula
\begin{equation}
\frac {\Delta E}{E}=(\gamma ^2-1)\frac {\Delta v}{v}  \label{ev}
\end{equation}
and the above statement is well supported. The velocity loss can become 
noticeable only for non-relativistic partons.

In what follows we consider very high energy processes. It is well known that
the gluons become the main component of the wave functions of the colliding
hadrons. At LHC, the $gg$-luminosity is at least by an order of magnitude
higher than the $\sum qq$-luminosity. Therefore the in-medium gluodynamics 
\cite{inmed} is considered below. It simplifies the formulae. Quarks can be 
easily included too \cite{dpri}.

The in-medium equations of gluodynamics 
differ from the in-vacuum equations by introducing a chromopermittivity
of the quark-gluon medium. Similar to the dielectric permittivity in
electrodynamics it describes the linear response of the matter to passing 
partons. 

Analogously to electrodynamics, one can treat the linear response
of the medium by the medium permittivity $\epsilon $
if $\bf E$ is replaced by ${\bf D} =\epsilon {\bf E}$ in $F^{\mu \nu}$.
 
Then, in terms of potentials the equations of {\it in-medium} gluodynamics are cast 
in the form \cite{inmed} (for $\epsilon $=const)
\begin{eqnarray}
\bigtriangleup {\bf A}_a-\epsilon \frac{\partial ^2{\bf A}_a}{\partial t^2}=
-{\bf j}_a -
gf_{abc}(\frac {1}{2} {\rm curl } [{\bf A}_b, {\bf A}_c]+
\epsilon \frac {\partial }
{\partial t}({\bf A}_b\Phi _c)+\frac {1}{2}[{\bf A}_b {\rm curl } {\bf A}_c]-  \nonumber \\
\epsilon \Phi _b\frac 
{\partial {\bf A}_c}{\partial t}- 
\epsilon \Phi _b {\rm grad } \Phi _c-\frac {1}{2} gf_{cmn}
[{\bf A}_b[{\bf A}_m{\bf A}_n]]+g\epsilon f_{cmn}\Phi _b{\bf A}_m\Phi _n), 
\hfill \label{f.6}
\end{eqnarray}

\begin{eqnarray}
\bigtriangleup \Phi _a-\epsilon \frac {\partial ^2 \Phi _a}
{\partial t^2}=-\frac {\rho _a}{\epsilon }+ 
gf_{abc}(-2{\bf A}_c {\rm grad }\Phi _b+{\bf A}_b
\frac {\partial {\bf A}_c}{\partial t}-\epsilon 
\frac {\partial \Phi _b}{\partial t}
\Phi _c)+  \nonumber  \\
g^2 f_{amn} f_{nlb} {\bf A}_m {\bf A}_l \Phi _b. \hfill  \label{f.7}
\end{eqnarray}
If one neglects the terms with explicitly shown charge $g$ in Eqs. (\ref{f.6}), 
(\ref{f.7}), one gets
the set of abelian equations, which differ from electrodynamical equations
by the color index $a$ only. The most important property of the solutions of 
these equations is that while the in-vacuum ($\epsilon = 1$) equations do not 
admit any radiation processes, for $\epsilon \neq 1$ there appear  
solutions of these equations with non-zero Poynting vector even in the classical 
approach. They predict such coherent effects as Cherenkov gluons, the wake and
transition radiation. This corresponds well to the 
microscopic description in which the matter, at early times after 
collision, is described in terms of the coherent classical field.

There are several classical polarization effects in the quark-gluon medium 
wherefrom one can determine its chromodynamical properties. Actually, this idea 
was first promoted in \cite{d1, d0}
relying on the similarity of quarks to electrons and gluons to photons.
The emission of Cherenkov gluons analogous to Cherenkov photons was predicted
and used for interpretation of a cosmic ray event where a ring-like structure 
was just observed \cite{apan} reminding Cherenkov findings of photon rings.
The strong support this idea got, however, 
much later from RHIC data on nucleus-nucleus collisions.

Other effects include the possible role of Cherenkov gluons in asymmetry
of shapes of resonances traversing the quark-gluon medium and the wake
radiation producing the shift in azimuthal distributions in non-central
collisions. I describe briefly only the ring effect referring to \cite{dl10}
for more details.  

This effect is derived from the classical solutions
of Eqs (\ref{f.6}), (\ref{f.7}) if the current with constant 
velocity ${\bf v}$ along the $z$-axis is considered:
\begin{equation}
\label{f.11}
{\bf j}({\bf r},t)={\bf v}\rho ({\bf r},t)=4\pi g{\bf v}
\delta({\bf r}-{\bf v}t).
\end{equation}

These solutions describe the cone-like emission of Cherenkov gluons at the 
typical angle
\begin{equation}
\label{f.10}
\cos \theta = \frac {1}{v\sqrt {\epsilon }}.
\end{equation}
It is constant for constant $\epsilon >1$.

Let us stress that this classical effect has collective non-perturbative origin 
even though $\alpha _S$ enters seemingly linearly. The
chromopermittivity $\epsilon $ hides the non-perturbative terms responsible
for the collective medium response. 

Similar to electrodynamics \cite{gr}, the 
energy-angular spectrum of emitted gluons \cite{dklv, wake} per the unit length
\begin{equation}
\frac {dN^{(1)}}{d\Omega d\omega }=\frac {\alpha _SC\sqrt x}{2\pi ^2}\left [
\frac {(1-x)\Gamma _t}{(x-x_0)^2+(\Gamma _t)^2/4}+\frac {\Gamma _l}{x}\right ], 
\label{9}
\end{equation}
where 
\begin{equation}
x=\cos ^2\theta, 
x_0=\epsilon_{1t}/\vert \epsilon _t \vert ^2v^2, \;\;\;                         
\Gamma _j=2\epsilon_{2j}/\vert \epsilon _j \vert ^2v^2, \;\;\;
\epsilon _j=\epsilon _{1j}+i\epsilon _{2j}.   \label{xep}
\end{equation}
The real ($\epsilon_1$) and imaginary ($\epsilon_2$) parts of $\epsilon $ are 
taken into account. 

It is clearly seen from Eq. (\ref{9})
that the transverse and longitudinal parts of the chromopermittivity are
responsible for the distinctly different effects. The ringlike Cherenkov
structure (conical emission) is exhibited in the 
first term of (\ref{9}). The second term defined by the longitudinal part 
of $\epsilon $ is in charge of the wake radiation. It is important that
both of them are independent on parton masses. This fact is supported by
experiment where there is no difference between energy losses of light and 
heavy quarks.

The first term was fitted \cite{dklv} to get the values of real ($\epsilon _1
\approx 6$) and imaginary ($\epsilon _2\approx 0.8)$ parts of the chromopermittivity. 
The fit to experimental data (with elliptic flow subtracted) is shown in Fig. 1 .

\begin{figure}[ht]
 \includegraphics[height=7cm, width=7cm]{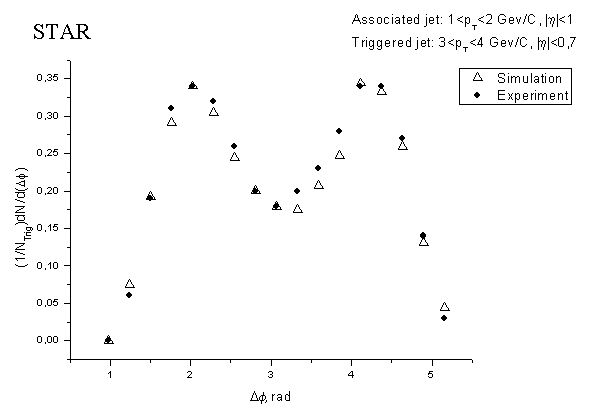} $\;$
\includegraphics[height=7cm, width=7cm]{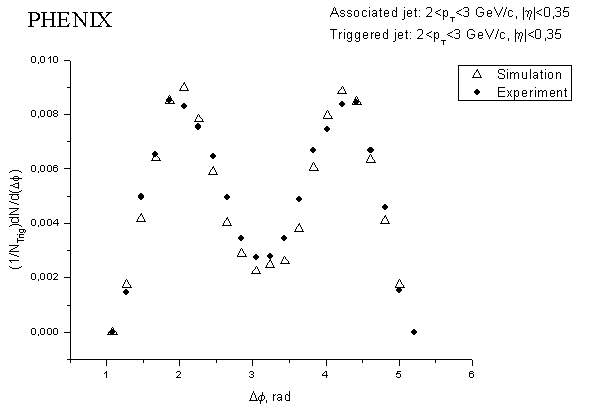}

 \caption{Associated azimuthal correlations at STAR and PHENIX: circles - experiment, triangles- theory.}
\end{figure}

These plots show the projection of rings on their diameter just as was done
in original papers by Cherenkov in 1937.
It is the very first determination of the chromopermittivity from experiment
in the definite range of the transverse momentum of created particles.
In principle, the chromopermittivity may depend on it (the well known dispersion
in electrodynamics). For discussion of this problem and other effects see
\cite{dl10}. 
Unfortunately, I had to omit many figures from original
presentation and refer the readers to it.

\section{Conclusions}

Let us briefly summarize what have we learned about the matter properties from 
comparison of experimental data with hydrodynamics and QCD.\\ 

{\bf HYDRODYNAMICS:}\\ 
temperature and energy density evolution, azimuthal (elliptic) flow, viscosity,
transfer properties, QGP lifetime, baryon density, hadrocomposition.\\

{\bf QCD:}\\
{\bf micro-approach} - jet quenching, the transfer coefficient $\hat q$ (LPM-effect),
QCD-vertex, role of angular ordering, color and medium size.\\
{\bf macro-approach} - chromopermittivity, parton density, energy loss to Cherenkov
gluons, free path length of partons, shape of resonances passing through the
medium, the wake, independence on quark masses.

\medskip

{\bf \large Acknowledgements}

\medskip

This work was supported by RFBR grants 09-02-00741; 08-02-91000-CERN and
by the RAN-CERN program.\\

\end{document}